\magnification=1200
\baselineskip=16truept

{\bf The central engine of gamma-ray bursters}
\bigskip
 W. Klu\'zniak{$^a$} \& M. Ruderman{$^b$}

\bigskip
{$^a$}University  of Wisconsin, Sterling Hall,
1150 University Ave., Madison, WI 53705

and Copernicus Astronomical Center, Warsaw

\smallskip
{$^b$}Columbia University, Pupin Hall, West 120th St., New York, NY 10027

and Istituto di Astrofisica Spaziale, via del Fosso de Cavaliere,
00133 Rome.

\bigskip
{\bf 
Gamma-ray bursts (GRBs) are thought to arise in relativistic blast-wave
shocks
at distances of $10^{15.5\pm 1.0}$cm from the
point where the explosive energy is initially released.$^{1,2}$ To
account for the observed duration and variability of the $\gamma$-ray emission
in most GRBs,
a central engine powering the shocks must
remain active for several seconds to many minutes but must strongly
fluctuate in its output on much shorter timescales.$^{2}$  We show
how a neutron star differentially rotating at millisecond 
periods (DROMP) could be such an engine. A magnetized DROMP
would repeatedly wind up
toroidal magnetic fields to about $10^{17}$G and only release
the corresponding magnetic energy, $E>10^{50}$erg, when each buoyant magnetic
field torus floats up to, and breaks through, the stellar surface.
The resulting rapid sub-bursts, separated by relatively quiescent
phases,
repeat until the kinetic energy of differential rotation is
exhausted by these events.  Calculated values of the energy released and of the
various timescales are in agreement with observations of GRBs. The amount
of matter ejected (baryon
loading) in each sub-burst may also be consistent with
theoretical requirements for a blast wave capable of giving
the X-ray, optical and radio
afterglows recently observed$^{3-7}$ from cosmological distances.
DROMPs could be created in
several kinds of astrophysical events; some of these
would be expected to occur at about the observed GRB rate of 
$\sim10^{-6}$/y per galaxy. The requisite very strong
internal differential rotation
could be imparted to neutron stars as they
are born or at the end of their existence: one consequence is that some DROMPs
may be created close to star forming regions while others may arise
far from galaxies.
}

\vfill\eject

Compelling evidence$^{3-7,8}$ now points to the cosmological origin of GRBs.
Observed fluences imply a typical total $\gamma$-ray energy release
$E_0\sim 10^{51}$--$10^{52}$erg for GRB sources at 3 Gpc.  The lack of
any high energy cutoff in the spectrum shows that the emission region
is optically thin to $\gamma$-rays above the pair creation
threshold---this is only possible if the effective radius, $d$, of
that region is hugely larger than the several light seconds travelled
by a photon in the GRB duration time ($t$), i.e. $d>>ct$. Therefore, the
immediate source of the $\gamma$-rays must move towards the observer
at relativistic speed with a Lorentz factor $\Gamma>>1$.

In contrast to the observed spectra of GRBs, which typically have a
power-law shape at high energies, photons emitted from a
``fireball'' with radiation in thermodynamic equilibrium with matter
would have a nearly thermal spectrum.$^{9,10}$
It appears necessary, therefore, for the
observed emission to arise in nonthermal processes in a region
sufficiently distant from ``ground zero'' of the explosion that the
expanding plasma has finally become optically thin to electron
scattering and pair creation by $\gamma$-rays.$^{2}$  The
relativistic blast-wave model, in which an external shock (and a
reverse shock) is formed as a
consequence of the significant slowing down of the expanding
relativistic ejecta (initial baryon mass $M_0$) after a mass $M_0/\Gamma$
has been swept up from the ambient medium, satisfies various
theoretical constraints for forming GRBs at cosmological distances.
The model has been especially successful in
explaining, and even predicting,$^{11-13}$ the long-lived X-ray, optical and
radio afterglows observed after some GRBs.

The detailed model of the blast wave
depends on both the energy released and the
density of the ambient medium. A match with
observations can be obtained for an explosion in an ambient density not
exceeding that of the interstellar matter (ISM) if the
initial kinetic energy, $E_0$, of the relativistic ejecta is of the order
of a tenth of a percent of the rest-mass energy of a neutron star,
i.e. $E_0\sim10^{51}$erg, and $\Gamma\ge100$. These values
suggest that a violent process
involving a neutron star, such as its merger with another comparably
compact object or its sudden creation as a millisecond pulsar, may power the
blast waves which give rise to a GRB.  However, even taking account of
relativistic expansion and beaming, the observed rapid variability of
flux would require a remnant central engine
which must remain active for time $t$, with the power output varying
on timescale up to $10^6$ times shorter (i.e., as short as a ms),
and then turn off.  The $\gamma-$ray emission could arise from internal
shocks$^{2,14,15}$,
as successive shells of relativistically moving plasma run into
each other.  The essential requirement is finding an engine
which emits such blasts with the correct time-scales for the energy
release, one in which the baryon
loading of the ejecta satisfies the necessary constraint
$M_0\le E_0\Gamma^{-1}c^{-2}\sim 10^{-5}M_\odot$ (for $\Gamma\sim100$),
and one for which all of this is not kept from being observed by an
opaque shell released in some initial blast.

The central engine of gamma-ray bursters must have the following properties
to account for the typical fluence in
each observed $\gamma-$ray sub-burst (``peak''),
for the number of peaks, $N_p$, the time
interval between peaks, $\tau$, and for the rapid rise times
and variability:
\par\noindent
a) an energy of $E_0\sim 10^{51}$erg must be released in each sub-burst;
\par\noindent
b) $N_p \sim 10$;
\par\noindent
c) between sub-bursts, the central engine should be dormant
for intervals $\tau\sim 1\,$s to $\sim 10^3$s;
\par\noindent
d) the engine should be capable of attaining its peak power within
milliseconds and of exhibiting large fluctuations thereafter.
\par\noindent
Further, to allow the formation of the relativistic shocks and the
ultimate emission of $\gamma$-rays at the distance $d$, discussed above,
\par\noindent
e) no more than $10^{-5}M_\odot$ of baryons may be carried in each
successive sub-burst.

We will argue that neutron stars with internal motions corresponding
to differential rotation with millisecond periods have at least four
of these properties and propose that they are the required
relatively long-lived central engines of GRBs. Moreover, such
neutron stars can be created in several kinds of astrophysical events at
rates approximating those of GRBs.

In the absence of interior magnetic fields, the differential
rotation would not be erased
in a very hot neutron star (e.g., by Ekman pumping) in less than a day.
However, magnetic field in the stellar interior could lead to a series of
explosive releases of the kinetic energy of differential rotation
until most of that energy is used up. The total available energy is then
$\hat I\Omega_d^2/2$, where $\Omega_d$ is the effective
difference in  angular frequency of
rotation and $\hat I$ is the corresponding
effective moment of inertia,
(which in a two-component model of differential rotation, is less than
 a quarter of the total stellar moment of inertia $\hat I<I/4$).
(For later convenience, we will write the
initial differential rotational frequency as
$\Omega_d=\Omega_4\times 10^4$s$^{-1}$, effective differential
moment of inertia as
$\hat I=\hat I_{44}\times 10^{44}$g cm$^2$, 
and initial interior field strength as $B_0=B_{12}\times 10^{12}$G.)

In a differentially rotating neutron star internal poloidal magnetic field
($B_0$) will be wound up into a toroidal configuration and amplified 
(to $B_\phi$) as one
part of the star (e.g. exterior) rotates about the other (e.g. core).
After $N_\phi$ revolutions,
$B_\phi=2\pi B_0 N_\phi$.
The toroidal field will be sufficiently buoyant to
overcome fully the (approximately radial)
stratification in neutron star composition only when a critical field
value, $B_f$, is reached.  Because neutron-star matter would be
brought up from the deep
interior to the stellar surface too quickly for weak interactions
to adjust its composition to the changing
ambient neutron to proton ratio, $y$, the difference in $y$ between
the interior and the subsurface layers results in a fractional
difference, $f$, between the density of
transported and ambient matter (at the same pressure $P$),
 $f=\rho^{-1}(\partial \rho/\partial y)_P\,\Delta y\sim2$\%. 
For hydrostatic equilibrium of a magnetic torus brought from the deep
interior to near the surface its magnetic buoyancy must be balanced by
the anti-buoyancy from its baryon ratio gradient. 
[We have neglected the effects of thermal gradients in this discussion].
Then,
$B_f^2/(8\pi)= f\rho (\partial P/\partial\rho)= f\rho c_s^2$, so that
$$B_f\approx 7\cdot10^{16}{\rm G}\times
 \left({\rho \over 10^{14}{\rm g\,cm^{-3}}}\right)^{1/2}, \eqno(1)$$
where we take the speed of sound $c_s\approx c/\sqrt10$.
The magnetic energy stored in a torus with this $B_f$ is
(for $\rho=3\cdot10^{14}$g/cm$^3$)
$$E_p\approx 6\cdot10^{50}(10\,V_B/V_*)\,{\rm erg}, \eqno(2)$$
Here $V_B/V_*$
is the fraction of the volume of the star occupied by the torus.
This $E_p$ is independent of the initial magnetic field and of the
initial differential rotational
period of the DROMP, as long as $\hat I\Omega_d^2/2>E_p$.

Only after the magnetic field reaches the critical value of eq. [1]
will the buoyant torus be able to
float up to break through the stellar surface.
The emergence of the torus
is accompanied by huge spin-down torques, reconnection of the new surface
magnetic
field and the quick release of an energy exceeding $E_p\sim 10^{51}$erg
(eq. [2]). This would be
a sub-burst, satisfying condition a). The rapidity
of the reconnection processes (occuring typically in $\sim10^{-4}$s,
the stellar radius divided by the
Alfv\'en speed)
would be expected to lead to exceedingly
short rise-times and large fluctuations of power, possibly in
agreement also with condition d).

The number of sub-bursts is the number of times the critical field
is built up and the magnetic toroid ejected. Then,
$$N_p={3\hat I\Omega_d^2\over B_f^2R^3}\left({V_*\over V_B}\right)
\approx6\Omega_4^2\hat I_{44}, \eqno (3)$$
in plausible agreement with condition (b) for typical values $\Omega_4\sim 1$,
$\hat I_{44}\sim 1$.
The interval between sub-bursts, i.e. the time to build up
the critical fields is
$$\tau={2\pi\over \Omega_d}{B_f\over B_0}\approx
 20\,{\rm s}\times B_{12}^{-1}\Omega_4^{-1}, \eqno(4)$$
in fair agreement with condition c). (Note the sensitive dependence on the
initial value of $B_0$.)

Eqs. (3) and (4) yield a total duration of
$t\approx 120\,{\rm s}\times B_{12}^{-1}\Omega_4$, after which
the (differential kinetic) energy stored in the central engine is
exhausted, or at least no longer capable of fully winding up another torus
so that it too can be released:
the gamma-ray burster turn off, apparently never to
be seen again.

We now turn to the expected birth rate of DROMPs,
keeping in mind the severe upper limits to the baryon
loading of the initial blast wave accompanying the turn-on of
the central engine. We are aware of four processes which
may give rise to DROMPs at a rate sufficient to
account for observed GRBs.

It has been argued$^{16,17}$ that the rate of coalescence of Hulse-Taylor type
neutron star binaries, $\sim10^{-6}/$y per galaxy, corresponds closely
to the observed GRB rate.According to Newtonian
calculations$^{18}$ the matter ejected in the merger is mostly confined to
the vicinity of the orbital plane, with no more than $10^{-5}M_\odot$ of
the baryons contained in the cone of opening half-angle $45^\circ$
around the rotation axis.
If the correct equation of state of dense matter is sufficiently
stiff---as recent observations of kHz quasi-periodic
oscillations in accreting neutron stars may
imply (ref. 19)---the post-merger core would not directly collapse to a black
hole. A massive neutron star rotating with a period of $P_0\sim1\,$ms
could be formed instead.
The same outcome would be achieved if the initial masses of the merging
neutron stars were very low, $M\le1M_\odot$.

Another process which might lead to the formation of massive ms pulsars at the
rate $\sim10^{-6}/$y per galaxy is accretion onto neutron stars in
low-mass X-ray binaries (LMXBs).  It seems plausible that at the end of
mass transfer many of the neutron stars are on the supramassive sequence,
i.e. they are supported by rapid rotation which delays a collapse to
a black hole possibly by as long as $10^{9}$y--$10^{10}$y (the pulsar
spin-down time). We would expect the onset
of collapse to initiate differential rotation, i.e. in the creation
of a short-lived DROMP and the ensuing GRB.

A third possibility is the rapid spin-up of a neutron star in a
catastrophic accretion event, perhaps resulting from a collision with
a white dwarf in a globular cluster.

Accretion induced collapse$^{20}$ (AIC) of certain evolved white dwarfs,
mainly$^{21}$ in 
globular clusters (GC), may be the ideal process in which a DROMP
giving rise to a GRB could be created. There is some evidence$^{22}$ that
a large fraction of the
ms pulsars in GCs have not been spun up in LMXBs and it has been
suggested that they are a product of AIC. If AIC occurs DROMPs are
nearly certain to result. Several AM Her type systems are known with
the white dwarf rotation rate of $\sim10^{-3}$s$^{-1}$ and magnetic fields
$\sim10^7$G, both values are expected to be amplified by a factor of
$\sim10^{6}$ in collapse to a neutron star, as pointed out by Usov$^{23}$ (who
suggested AIC creation of ms pulsars with a $10^{15}$G dipole field
which could directly energize the blast-wave with their spin-down
power, thus giving rise to a single-peaked GRB of $\sim30\,$s duration). 
The critical question is whether an initial blast
wave associated with the AIC formation of the strongly magnetized ms pulsar
has sufficiently small mass to allow the subsequent GRB phase to be observable.
On the other hand,
if that blast carries away considerable mass$^{24}$, the remnant neutron star
mass would be much less than $1.4M_\odot$. Then, in the dense cores of GCs,
Hulse-Taylor-like neutron star binaries might be formed with much lighter
neutron stars, whose subsequent merger would almost certainly
lead to the formation of  DROMPs.

We thank Drs. J. Katz, S. Colgate, M. Rees, J.R. Wilson
and G. Bisnovatyi-Kogan, for informative conversations and the Aspen Center
for Physics for its hospitality.
A torus of neutron star matter has been suggested as a possible remnant of
a Hulse-Taylor merger.$^{25}$ Katz$^{26}$ has shown how its differential
rotation
could build up a huge toroidal magnetic field in it. In
general-relativistic calculations$^{27}$   of the final stages of evolution of
neutron-star binaries, Wilson and collaborators
find internal motions amplifying magnetic
fields to $10^{17}$G.
Bisnovatyi-Kogan
has suggested a single huge ejection of stellar matter by a differentially
rotating stellar core toroidal magnetic field as a Type~I supernova
explosion mechanism.

\vfill\eject

\par\noindent
REFERENCES

\par\noindent
~1. M\'eszar\'os, P. \& Rees. M. 1993, Astrophys. J.
 405, 278--284.

\par\noindent
~2. Sari, R. \& Piran, T. 1997, Astrophys. J.
 485, 270--273.

\par\noindent
~3. van Paradijs, J. {\it et al.} 1997, Nature 386, 686--689.

\par\noindent
~4. Galama, T.J. {\it et al.} 1997, IAU Circ. 6655.

\par\noindent
~5. Djorgovski, S.G. {\it et al.} 1997, Nature 387, 876--878.

\par\noindent
~6. Frail, D.A. {\it et al.} 1997, IAU Circ. 6662.

\par\noindent
~7. Fruchter, A. {\it et al.} 1997, IAU Circ. 6674.

\par\noindent
~8. Fishman, G.J. \& Meegan, C.A. 1995, Ann. Rev. Astron. Astroph. 33,
    415--458.

\par\noindent
~9. Paczy\'nski, B. 1986, Astrophys. J.
 308, L43--L46.

\par\noindent
10. Goodman, J. 1986, Astrophys. J.
 308, L47--L50.

\par\noindent
11. Paczy\'nski, B. \& Rhoads, J.E. 1993, Astrophys. J.
 418, L5--L8.

\par\noindent
12. M\'eszar\'os, P. \& Rees. M. 1997, Astrophys. J.
 476, 232--237.

\par\noindent
13. Vietri, M. 1997, Astrophys. J.
 478, L9--L12.
 
\par\noindent
14. M\'eszar\'os, P. \& Rees. M. 1994, Mon. Not. R. astron. Soc.,
    269, L41--L43.

\par\noindent
15. Mochkovitch, R. \& Daigne, F. 1997, in 4th Gamma-Ray Burst Symposium,
    C.A.~Meegan {\it et al.}, eds., (New York: AIP), in press.

\par\noindent
16. Paczy\'nski, B. 1991, Acta Astron. 41, 257--267.

\par\noindent
17. Narayan, R., Piran, T. \& Shemi, A. 1991, Astrophys. J.
 379, L17--L20.

\par\noindent
18. Rasio, F. \& Shapiro, S.L. 1992, Astrophys. J.
 401, 226--245.

\par\noindent
19. Klu\'zniak, W. 1997, preprint, astro-ph/9712243.

\par\noindent
20. Nomoto, K. 1991, Astrophys. J. 367, L19--L22.

\par\noindent
21. Bailyn, C. \& Grindlay, J. 1990, Astrophys. J. 353, 159--167.

\par\noindent
22. Chen, K., Middleditch, J. \& Ruderman, M. 1993, Astrophys. J.
 408, L17--L20.

\par\noindent
23. Usov, V.V. 1992, Nature, 357, 472--474.

\par\noindent
24. Colgate, S., Fryer, C.L. \& Hand, K.P. 1997, in
    Thermonuclear Supernovae, P. Ruiz-Lapuente {\it et al.}, eds.,
    (Dordrecht: Kluwer), pp. 273--302.

\par\noindent
25. Narayan, R., Paczy\'nski, B. \& Piran, T. 1992, Astrophys. J.
 395, L83--L86.

\par\noindent
26. Katz, J. 1997, Astrophys. J. 490, 633--641.

\par\noindent
27. Mathews, G.J., Wilson, J.R. \& Maronetti, P. 1997, Astrophys. J.
 482, 929--941.

\bigskip\par\noindent
This work supported in part by NASA and Poland's Committee for Scientific
Research (KBN).

\vskip1truein
\par\noindent
FIGURE CAPTION

\par\noindent
{\bf Fig. 1} The magnetic torus emerging from the neutron star.

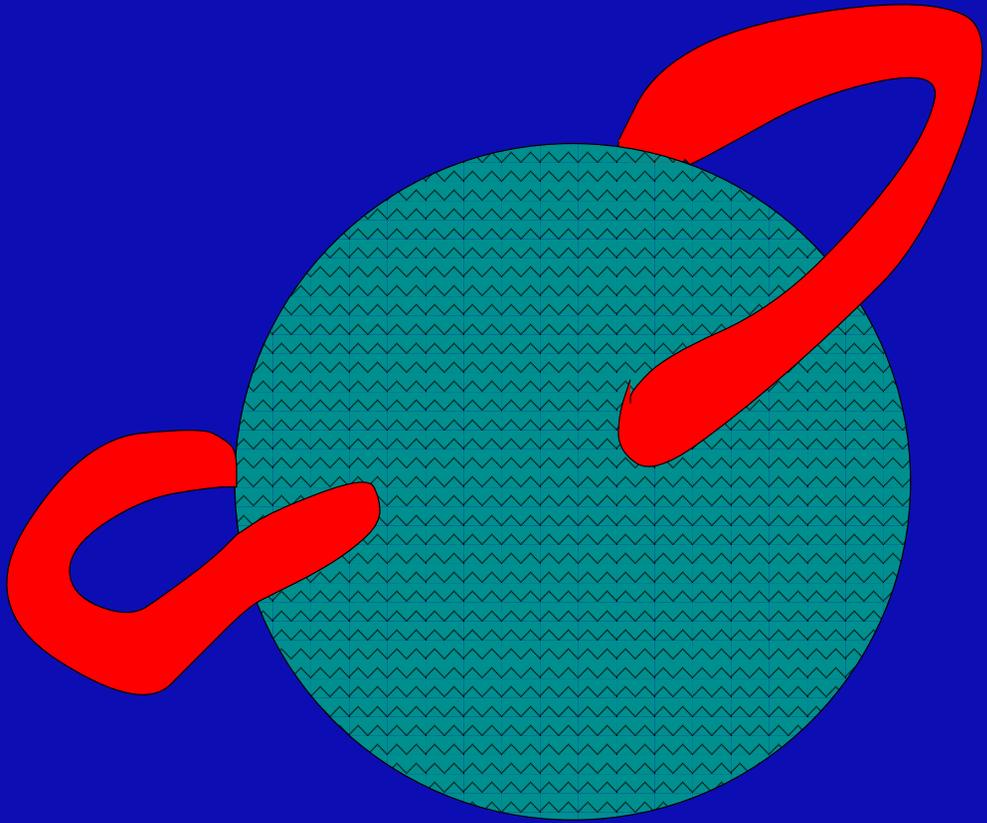
\bye